\documentstyle[12pt]{article}

\newcommand{\comment}[1]{}

\newcommand{\be}{\begin{equation}}
\newcommand{\ee}{\end{equation}}
\newcommand{\lcal}{{\cal L}}

\newcommand{\BW}{{\mbox{BW}}}

\begin{document}
\everymath={\displaystyle}
\vspace*{-2mm}
\thispagestyle{empty}
\noindent
\hfill TTP93--28\\
\mbox{}
\hfill  October 1993  \\
\vspace{0.5cm}
\begin{center}
  \begin{Large}
  \begin{bf}
   PSEUDOSCALAR MASS EFFECTS  IN DECAYS
   OF TAUS  WITH THREE PSEUDOSCALAR MESONS
   \\
  \end{bf}
  \end{Large}
  \vspace{0.8cm}
  \begin{large}
   Roger Decker, Markus Finkemeier and Erwin Mirkes\footnote{
Address after Sept.\ 1, 1993:
Physics Dept., University of Wisconsin, Madison WI 53706, USA
} \\[5mm]
    Institut f\"ur Theoretische Teilchenphysik\\
    Universit\"at Karlsruhe\\
    Kaiserstr. 12,    Postfach 6980\\[2mm]
    76128 Karlsruhe\\ Germany\\
  \end{large}
  \vspace{4.5cm}
  {\bf Abstract}
\end{center}
\begin{quotation}
\noindent
%
{We study the effects of non-vanishing pseudoscalar masses in
$\tau$ decays into three mesons.
The hadronic matrix elements are obtained by using
the generalized structure of the chiral currents with nonvanishing pseudoscalar
masses and implementing the low-lying resonances in the different channels.
We demonstrate that suitable angular distributions are sensitive to the mass
 effects
in the chiral Langrangian. Numerical results for the rele\-%
vant structure functions are
given for the decay modes $\tau\rightarrow \nu\pi^-\pi^-\pi^+$,
$\nu K^-\pi^-\pi^+$ and $\nu K^-\pi^-K^+$.
}

\end{quotation}
\newpage

%
%
\section{Introduction}
In this paper we investigate the effects of the non-vanishing
pseudoscalar meson masses on predictions which use chiral symmetry for
calculating tau decays into neutrino and three pseudoscalar mesons
(pions and/or kaons).
The usual
philosophy employed to derive  the hadronic matrix elements  is to introduce
resonances in the different  channels with constant couplings.
These couplings are then fixed by the low energy theorems which follow from the
spontaneously broken chiral symmetry of  QCD
and can be calculated using
effective chiral Lagrangians \cite{ttp9225,DMREV,KM1,KM2,diverse}.
But note that all these authors use the chiral Lagrangian
for exactly massless pseudoscalar mesons,
which results in a completely transverse (ie.\ conserved)
hadronic amplitude with no spin-0 contribution.
All what is then known about the
size of the spin-0 amplitude is
the PCAC (partial conserved axial current) argument that
the spin-0 amplitude should be suppressed by a factor $m_\pi^2/Q^2$
relative to the spin-1 amplitude ($Q^2$ is the invariant mass
squared of the hadronic system).
The chiral Lagrangian for massive
pseudoscalar mesons, however, is also well known \cite{massive}
and so it would seem to be the best
approach to start from the generalized chiral limit which it describes
and then include the resonances. This is what we do in this present paper.
Of course the hadronic current predicted
by such a model is no longer transverse and results in a nonvanishing spin-0
amplitude.

Now the question arises whether and how the effects of the spin-0 amplitude can
be measured.
In many experimental analyses  of $\tau \to 3 \pi \nu_\tau$ decays, up to now
the spin-0 part of the hadronic system has been neglected.
This was justified by the above mentioned PCAC suppression of this scalar part
relative to the spin-1 part.
Assuming the mean $Q^2$ to be about $m_{a_1}^2$ in the decay
$\tau \to 3 \pi \nu_\tau$, the spin-0 amplitude
is expected to be about  $1 \% $ of that of the spin-1 amplitude
and therefore the relative contribution to the total
rate for $\tau \to 3 \pi \nu $ to be of the order of $10^{-4}$. For decays
modes with kaons the corresponding quantity $m_K^2/Q^2$ could be as
large as $10 \%$, but even then the contribution to the total width would not
be more than a few percent.
Arguments along these lines show that the scalar contributions are not
measurable in the total decay width. But of course these are only
very crude order of magnitude arguments which could be modified by
higher-lying resonances. However, an inspection of the
effects of the $\pi (1300)$ has shown no important enhancement \cite{ttp9225}.

Therefore we must consider  other quantities for
a measurement of the scalar form factor. It appears that
angular distributions \cite{KM1,KM2}
are sensitive to the interference of spin-0 and spin-1 form factors
and could make the measurement of the spin-0 form factor in high
statistic experiments possible.
We will therefore study in this paper if this is actually feasible.

This paper is organized as follows: In Sec.\ 2 we review
the kinematics, the general form of the hadronic current and
the definitions of the angular moments. In Sec.\ 3 we present the generalized
chiral predictions for all the ten possible decay channels with pions
and kaons in the final state. In Sec.\ 4 we implement vector and
axial-vector resonances, give
general formulae for the form factors and the specialization to the three
most interesting modes, viz. to $\tau \to$ either $\pi^- \pi^- \pi^+$ or
$K^- \pi^- \pi^+$ or $K^- \pi^- K^+$. The numerical results for these
three channels are presented in Sec.\ 5, and in Sec.\ 6 we draw our
conclusions.
%

\section{Hadronic Current and Angular Distributions}
%
We consider the decay
\be
   \tau \to M_1(k_1) M_2(k_2) M_3(k_3) \nu_\tau
\ee
The hadronic matrix element involved can be parametrized in terms of
four form factors:
\begin{eqnarray}
   J^\mu & = &\langle M_1(k_1) M_2(k_2) M_3(k_3) |( V^\mu-A^\mu) | 0
\rangle
\nonumber \\
   & = & \Bigg[F_1(Q^2,s_1,s_2) (k_1 - k_3)_\nu
       + F_2(Q^2,s_1,s_2) (k_2 - k_3)_\nu \Bigg]
   \Bigg[g^{\mu\nu} - \frac{Q^\mu Q^\nu}{Q^2} \Bigg]
\nonumber \\ & &
   + F_3(Q^2,s_1,s_2) \epsilon^{\mu \alpha \beta \gamma}
     {k_1}_\alpha {k_2}_\beta {k_3}_\gamma
   + F_S(Q^2,s_1,s_2) Q^\mu
\end{eqnarray}
where
$V^\mu$ and $A^\mu$ are the vector and axialvector quark currents,
respectively,
\be
    Q = k_1 + k_2 + k_3
\ee
and the invariant masses $s_j$ are defined by
\begin{eqnarray}
   s_1  & = & (k_2 + k_3)^2 \nonumber\\
   s_2  & = & (k_1 + k_3)^2
\end{eqnarray}
The differential decay rate  is obtained from
  \begin{equation}
d\Gamma(\tau\rightarrow \nu_\tau\,3h)=
\frac{G^{2}}{4m_\tau}\,\bigl(^{\cos^2\theta_{c}}_{\sin^2\theta_{c}}\bigr)\,
\left\{L_{\mu\nu}H^{\mu\nu}\right\}
\,d\mbox{PS}^{(4)}
 \label{decay}
  \end{equation}
where $L_{\mu\nu}=M_\mu (M_\nu)^\dagger$, $M^\mu$ is the leptonic current and
$H^{\mu\nu} =  J^{\mu}(J^{\nu})^{\dagger}$.

The considered decays  are most easily analyzed in the
hadronic rest frame
$\vec{k}_{1}+\vec{k}_{2}+\vec{k}_{3}=\vec{Q}=0$.
The orientation of the hadronic
system is characterized by
three Euler angles ($\alpha,\beta$ and $\gamma$) as introduced in
\cite{KM1,KM2}.

Note that in current $e^++e^-$ ($\to \tau^+\tau^-(\to \nu_\tau 3$ mesons))
experiments only two out of the three Euler angles are measurable.
The reason for this is that the rest frame of the $\tau$ can not be
reconstructed or equivalently the direction of flight of the
$\tau$ in the hadronic rest frame is not known. Instead the
direction of the Labframe in hadronic rest frame is introduced.
The measurable Euler angles are defined by
 \begin{eqnarray}
\cos\beta&=&\hat{n}_{L}\cdot\hat{n}_{\perp} \label{cbetadef}\\[2mm]
\cos\gamma&=&-\frac{\hat{n}_{L}\cdot\hat{k}_{3}}{
             |\hat{n}_{L}\times\hat{n}_{\perp}|}\label{cgammadef}
 \end{eqnarray}
where ($\hat a$ denotes a unit three-vector)
 \begin{itemize}
\item $n_L$ is the direction of the labframe in the hadronic restframe.
Note that  $\hat n_L=-\hat n_Q$, with $\hat n_Q$  the direction of the
hadrons in the labframe,
\item
 $\hat{n}_\perp =\hat k_1\times \hat k_2$, the normal to the plane
defined by the momenta of particles 1 and 2.
 \end{itemize}

Performing the integration over the
momentum of the unobserved neutrino and the Euler angle $\alpha$
we obtain the differential decay width for a polarized $\tau$
\cite{KM1,KM2}:
  \begin{eqnarray}
d\Gamma(\tau\rightarrow 3h)&=&
           \frac{G^{2}}{2m_\tau} \,
\bigl(^{\cos^2\theta_{c}}_{\sin^2\theta_{c}}
\bigr)
\, \left\{\sum_{X}\bar{L}_{X}W_{X}\right\}\times \label{diffrat}\\[3mm]
&&\frac{1}{(2\pi)^{5}}\frac{1}{64}
\frac{(m_\tau^{2}-Q^{2})^{2}}{m_\tau^{2}}\,
     \frac{dQ^{2}}{Q^{2}}\,ds_{1}\,ds_{2}
     \,\frac{d\gamma}{2\pi}\,
      \frac{d\cos\beta}{2}\,    \frac{d\cos\theta}{2}
      \nonumber
\end{eqnarray}
The main advantage of working in the hadronic rest frame is that the
product $L_{\mu\nu}H^{\mu\nu}$ reduces to a sum of 16 hadronic
structure functions $(W_X)$
\begin{equation}
L_{\mu\nu}H^{\mu\nu}\to\sum_X \bar{L}_XW_X
\end{equation}
The angle $\theta $, a kinematical angle,  can be seen as the angle between the
unmeasured direction of
flight of the $\tau$ in the labframe and $\hat{Q}$ in the $\tau$
restframe, and is
obtained from the hadronic energy in the labframe
$E_h$ by \cite{KM1,KM2}
 \begin{equation}
\cos\theta = \frac{\left(2xm_\tau^{2}-m_\tau^{2}-Q^{2}\right)}{
          (m_\tau^{2}-Q^{2}) \sqrt{1-4m_\tau^{2}/s}}
\label{cthdef}
  \end{equation}
with
  \begin{equation}
\hspace{0.1cm}  x  = 2\frac{E_{h}}{\sqrt{s}}
\hspace{1cm}  s = 4 E^{2}_{{beam}}
  \end{equation}
Finally, the angle between the unmeasured tau direction and
that of the lab in the hadron restframe ($\vec Q=0$) is needed
(for $\overline{K_i}$ in Eqn.\ (\ref{ks}) and it
can  also computed from the energy $E_h$:
  \begin{eqnarray}
\cos\psi &=&
       \frac{x(m_\tau^{2}+Q^{2})-2Q^{2}}{(m_\tau^{2}-Q^{2})
       \sqrt{x^{2}-4Q^{2}/s}}
\label{psi}
  \end{eqnarray}

Different combinations of the hadronic form factors can be measured by
considering moments
\begin{equation}
\langle f(\beta ,\gamma)\rangle \propto \int f(\beta ,\gamma)
 \sum_X L_{X}H^{X}d\cos\beta d\gamma
\end{equation}
which have been  defined in \cite{KM2}.

The simplest moment ($f(\beta ,\gamma)=1$)
is proportional to the angular integrated rate:
\begin{equation}
\langle 1\rangle \propto
(2 K_1+3K_2)(W_A+W_B)+3K_2W_{SA}
\end{equation}
where ${K_i}(\theta,\psi, P)$ are known functions
of kinematical variables $\theta$ and $\psi$ and the
$\tau$ polarisation $P$ \cite{KM2}.
The hadronic structure functions $W_{X}$ depends on $Q^{2},s_{1}$ and $s_{2}$.
$W_{SA}$ and $W_{A}+W_{B}$ are
 closely related to the spin-0 and spin-1
part of the spectral functions:
  \begin{eqnarray}
\label{eqnrate1}
\rho_{0}(Q^{2})&=&\frac{1}{2}\frac{1}{(4\pi)^{4}}\frac{1}{Q^{4}}\,
                \int ds_{1}\,ds_{2}\,\,W_{SA}\label{st0}\\[2mm]
\rho_{1}(Q^{2})&=&\frac{1}{6}\frac{1}{(4\pi)^{4}}\frac{1}{Q^{4}}\,
                \int ds_{1}\,ds_{2}\,\,(W_{A}+W_{B})\label{st1}
  \end{eqnarray}
and
  \begin{equation}
\label{eqnrate2}
\hspace{-1cm}
\Gamma(\tau\rightarrow 3h)=
           \frac{G^{2}}{4m_{\tau}}
(g_{V}^{2}+g_{A}^{2})
\bigl(^{\cos^2\theta_{c}}_{\sin^2\theta_{c}}\bigr)
\frac{1}{(4\pi)}
\int dQ^{2} \,(m_{\tau}^{2}-Q^{2})^{2}
\left\{\rho_{0}+\left(1+\frac{2Q^{2}}{m_{\tau}^{2}}\right)\rho_{1} \right\}
      \nonumber
\end{equation}
Note that
that the spin-$0$ contribution is very small as
compared to the spin-$1$ part \cite {ttp9225}.

In order to measure spin-$0$ contributions we have to consider
interference of spin-$0$ and  spin-$1$ terms.
Therefore  the following moments are of interest
\begin{eqnarray}
f(\beta ,\gamma)&=& \cos\beta\nonumber\\
f(\beta ,\gamma)&=& \sin\beta\cos\gamma\nonumber\\
f(\beta ,\gamma)&=& \sin\beta\sin\gamma
\end{eqnarray}
In the notation of \cite{KM2} these moments are proportional to
\begin{eqnarray}
\label{ks}
\langle  \cos\beta\nonumber\rangle&
\propto &\overline{K_3}(\theta,\psi, P)\, W_E-\overline{K_2}(\theta,\psi, P)
\,W_{SF}  \\
\langle \sin\beta\sin\gamma\rangle
&\propto& -\overline{K_3}(\theta,\psi, P)\, W_G-
\overline{K_2}(\theta,\psi, P)\, W_{SD} \nonumber\\
\langle \sin\beta\cos\gamma \rangle
&\propto& -\overline{K_3}(\theta,\psi, P)\, W_I
+\overline{K_2}(\theta,\psi, P)\, W_{SB}
\label{moments}
\end{eqnarray}
where $\overline{K_i}(\theta,\psi, P)$ are known functions
of kinematical variables $\theta$ and $\psi$ and the
$\tau$ polarisation $P$ \cite{KM2}.

The needed hadronic structure functions $W_X$ are related to
the hadronic form factors.
Let us consider the
hadronic restframe with $z$ and
$x$  axis are aligned with $\vec{n}_{\perp}$ and
 ${\vec{k}_{3}}/{|\vec{k}_{3}|}$, respectively.
In this frame the momenta of the hadrons are given as follows:
  \begin{eqnarray}
k_{3}^{\mu}&=&(E_{3},k_{3}^{x},0,0)\nonumber\\
k_{2}^{\mu}&=&(E_{2},k_{2}^{x},k_{2}^{y},0)\label{qmomenta}\\
k_{1}^{\mu}&=&(E_{1},k_{1}^{x},-k_{2}^{y},0)\nonumber
  \end{eqnarray}
Then the  following variables are useful to express
the hadronic structure functions $W_{X}$
   \begin{eqnarray}
x_{1}&=& k_{1}^{x}-k_{3}^{x}\nonumber\\[1mm]
x_{2}&=& k_{2}^{x}-k_{3}^{x}\nonumber\\[1mm]
x_{3}&=& k_{1}^{y}=-k_{2}^{y}\\[1mm]
x_{4}&=& =\sqrt{Q^{2}}x_{3}k_{3}^{x}\nonumber
   \end{eqnarray}
Using these variables the following results hold
  \begin{eqnarray} \label{ws} \hspace{3mm}
W_{A}  &=&   \hspace{3mm}(x_{1}^{2}+x_{3}^{2})\,|F_{1}|^{2}
           +(x_{2}^{2}+x_{3}^{2})\,|F_{2}|^{2}
           +2(x_{1}x_{2}-x_{3}^{2})\,\mbox{Re}\left(F_{1}F^{\ast}_{2}\right)
                                   \nonumber \\[5mm]
W_{B}  &=& \hspace{3mm} x_{4}^{2}|F_{3}|^{2}
                                   \nonumber \\[5mm]
W_{SA} &=& \hspace{3mm} Q^{2}\,|F_{4}|^{2}\nonumber\\[5mm]
W_{E}  &=& -2x_{3}(x_{1}+x_{2})\,\mbox{Im}\left(F_{1}
                    F^{\ast}_{2} \right) \label{walldef}\nonumber\\[5mm]
W_{G}  &=&- 2x_{4}\left[x_{1}\,\mbox{Re}\left(F_{1}F^{\ast}_{3}\right)
                     + x_{2}\,\mbox{Re}\left(F_{2}F^{\ast}_{3}\right)\right]
                                   \nonumber \\[5mm]
W_{I}  &=&- 2x_{3}x_{4}\left[\,\mbox{Re}\left(F_{1}F^{\ast}_{3}\right)
                     -\,\mbox{Re}\left(F_{2}F^{\ast}_{3}\right)\right]
                                   \nonumber  \\[5mm]
W_{SB}  &=&  2\sqrt{Q^2}\left[\,x_1\mbox{Re}\left(F_{1}F^{\ast}_{4}\right)
                     +\,x_2\mbox{Re}\left(F_{2}F^{\ast}_{4}\right)\right]
                                   \nonumber \\[5mm]
W_{SD}  &=&
      2x_3\left[\,\mbox{Re}\left(F_{1}F^{\ast}_{4}\right)
                     -\,\mbox{Re}\left(F_{2}F^{\ast}_{4}\right)\right]
                                   \nonumber \\[5mm]
W_{SF}  &=&
     -2\sqrt{Q^2}x_4\, \mbox{Im}\left(F_{3}F^{\ast}_{4}\right)
\end{eqnarray}
Before  discussing our model we would like to remind the reader some
properties of the structure functions
\begin{itemize}
\item Let us consider the case of the decay into three pions.
Due to $G$-parity conservation the form factor $F_3$ vanishes.
Using Eqn.\ (\ref {ws}) we observe that $W_G$ and $W_I$ are vanishing and
therefore in this case (three pions) a nonvanishing
$\langle \sin\beta\cos\gamma\rangle$ and or $\langle \sin\beta\sin\gamma
\rangle$ yields a clean signature of the presence a scalar
contribution.
\item $\langle \cos\beta\rangle$ yields a measurement of the parity
violation in $\tau$ decay and for pions $W_{SF}=0$.
\item In general a measurement of the scalar parts is only possible if
the spin-0 structure functions $W_{SA,SB,SD,SF}$ in Eqn.(\ref{ws})
 are comparable with the spin-1 structure functions at least in some
kinematical
areas.
\end{itemize}
\section{The Chiral Limit}
The generalized chiral limit, ie.\ with nonvanishing pseudoscalar masses,
 is most conveniently described by the effective Lagrangian \cite{massive}
\be
   \lcal^{(2)} = \frac{f_\pi^2}{4}\mbox{tr}
       (\partial_\mu U \partial^\mu U^\dagger)
	+ \frac{f_\pi^2 \mu}{2} \mbox{tr}(M U^\dagger + U^\dagger M)
\ee
where U is the exponential of the pseudoscalar fields,
\be
   U = \exp\left\{
       \frac{\sqrt{2}i}{f_\pi} \left(
  	\begin{array}{ccc}
	\pi^0/\sqrt{2} + \eta/\sqrt{6} & \pi^+ & K^+ \\
	\pi- & - \pi^0/\sqrt{2}+\eta/\sqrt{6} & K^0 \\
	K^- & \bar{K}^0 & - 2 \eta/\sqrt{6}
     	\end{array}
       \right)
       \right\}
\ee
$M$ the quark mass matrix
\be
   M = \mbox{diag}(m_u, m_d, m_s)
\ee
and
\be
   f_\pi = 93.3 \mbox{MeV}.
\ee
 From this Lagrangian the matrix element $H_{chiral}^\mu$ of the
axial weak hadronic current $A^\mu$ between the hadronic vacuum and
a state with three pseudoscalar mesons $M_1$, \dots $M_3$ is derived as
\begin{eqnarray}
   H_{chiral}^\mu 
   = \frac{2 \sqrt{2}}{3 f_\pi} A^{(123)}
   & \Big\{ & \left[ (k_1 - k_3)^\mu - \frac{1}{2}
     \frac{(q+k_2)\cdot(k_1-k_3)}{q^2 - m_{(123)}^2} q^\mu \right]
\nonumber \\
 & &    + G^{(123)}\left[ (k_2 - k_3)^\mu - \frac{1}{2}
     \frac{(q+k_1)\cdot(k_2-k_3)}{Q^2 - m_{(123)}^2} q^\mu \right]
 \nonumber \\ & &
 + \frac{1}{2} \frac{X^{(123)}}{Q^2 - m_{(123)}^2} q^\mu\Big\}
\label{eqn5}
\end{eqnarray}
The values for $A^{(123)}$, $G^{(123)}$, $m_{(123)}$ and $X^{(123)}$
for the respective channels are found in Tab.\ \ref{tab1}.
\begin{table}
\caption{Parameters for the respective three meson channels
(The charge conjugated channels are obtained by reversing
the sign of $A^{(123)}$.)
}
\label{tab1}
$$ \begin{array}{|c|c|c|c|c|c|}
M_1 M_2 M_3 &  A^{(123)} & G^{(123)} & m_{(123)}^2
& X^{(123)} \\
\hline \hline
\pi^- \pi^- \pi^+ &  \cos \theta_c & 1 & m_\pi^2 & 2 m_\pi^2 \\
\pi^0 \pi^0 \pi^- &  \cos \theta_c & 1 & m_\pi^2 & m_\pi^2 \\
K^- \pi^- K^+   & - 1/2 \cos \theta_c & 1 & m_\pi^2 & m_\pi^2 + m_K^2 \\
{K}^0 \pi^- \bar{K}^0 &- 1/2 \cos \theta_c &1& m_\pi^2& m_\pi^2 + m_K^2 \\
K^- \pi^0 {K}^0 &  3/(2 \sqrt{2}) \cos \theta_c & 0 & m_\pi^2 & 0 \\
\pi^0 \pi^0 K^- & 1/4 \sin \theta_c & 1 & m_K^2 & - 2 (m_\pi^2 + m_K^2) \\
K^- \pi^- \pi^+ & - 1/2 \sin \theta_c & 1 & m_K^2 & m_\pi^2 + m_K^2 \\
\pi^- \bar{K}^0 \pi^0 & 3/(2 \sqrt{2}) \sin \theta_c  & 0 & m_K^2 & 0 \\
K^- K^- K^+ &  \sin \theta_c & 1 & m_K^2 & 2 m_K^2 \\
K^- \bar{K}^0 {K}^0 & - 1/2 \sin \theta_c & 1 & m_K^2 & 2 m_K^2
\\ \hline \hline
\end{array}
$$
\end{table}
Note that in the strict chiral limit, ie.\ when $m_u = m_d = m_s = 0$,
the matrix element,
which in this limit we will denote by $H_{chiral,0}^\mu$,
 becomes transverse, because then $m_{(123)}$ and
$X^{(123)} = 0$ and $1/2 (q+k_2)\cdot (k_1 - k_3) = q \cdot (k_1 - k_2)$ (and
similarly for $(1 \leftrightarrow 2)$.

\section{Implementation of Resonances}
The chiral current of the last section is the $O(P^2)$ theorem of low energy
QCD and can only be expected to be correct for very small momentum transfers
(very small compared with typical resonances masses such as $m_\rho^2$, say).
For larger momenta, effects of the order of $O(P^4)$, $O(P^6)$,~\dots must
successively be taken into account, and when the momentum transfers can
actually become larger than the resonance masses, all orders in $O(P^2)$ must
be summed up. The leading effect of this series may be described
by Breit-Wigner resonances $\BW(s)$:
\be
    BW_X(s) = \frac{m_X^2}{m_X^2 - i m_X \Gamma_X(s) - s} =
    \sum_{n=0}^{\infty}\left(\frac{s+im_X \Gamma_X(s)}{m_X^2}\right)^n
\ee
And so the usual approach to extrapolate to higher momenta is to write down a
current including Breit Wigners describing possible final state resonances
in such a way that in the low energy limit, where all terms of higher order
than $O(P^2)$ are neglected,  the current reduces to the
correct chiral limit. Note that in the chiral counting
not only the pseudoscalar momenta $k_1$, $k_2$ and $k_3$, but also
the pseudoscalar masses $m_\pi$, $m_K$ and
the coupling to the
external gauge field of the $W$ count as $O(P)$.
And so taking the limit of the order $O(P^2)$ for the amplitude means
taking the limit of the order $O(P^0)$ for the form factors. It turns out
that this limit is always obtained from our formulae by putting the involved
Breit-Wigner resonance factors equal to one.
Let us consider the three pion case first for simplicity. The vector meson
dominance model of Fig.\ 1 gives the following current:
\begin{eqnarray}
    H^\mu & = & C^{(3\pi)} \Bigg\{ BW_{a_1}(Q^2)\; \left(g^{\mu\nu} -
    \frac{q^\mu q^\nu}{m_{a_1}^2} \right) \;
    \Gamma_{\nu \alpha}
    BW_{\rho}(s_2)\;
\nonumber \\
  & & \times \left(g^{\alpha \beta} -
   \frac{(k_1 + k_3)^\alpha (k_1 + k_3)^\beta}
   {m_\rho^2} \right) \;
   (k_1 - k_3)_\beta + (1 \leftrightarrow 2) \Bigg\}
\end{eqnarray}
Here $C^{(3\pi)}$ is an overall factor (product of couplings and the like).
$\Gamma_{\mu \nu}$ is proportional to the vertex describing the
coupling of the $a_1$ to the $\rho \pi$. The most general form for
$\Gamma_{\mu \nu}$ for off-shell particles
contains five form factors. The ansatz
in Ref. \cite{ttp9225} corresponds to the transverse form
\be
  \Gamma_{\mu\nu} \Big( a_1(q_\mu) \to \rho(k_\nu) \pi \Big)
  = g_{\mu\nu}
    - \frac{q_\mu q_\nu}{Q^2}
\ee

With this ansatz for $\Gamma_{\mu\nu}$ and the normalization
\be
     C^{(3\pi)} = \frac{2 \sqrt{2}}{3 f_\pi} A^{(3\pi)}
   =  \cos \theta_C \frac{2 \sqrt{2}}{3 f_\pi}
\ee
the hadronic current $H^\mu$ reduces to the strict chiral limit of
vanishing quark masses, ie.\ the transverse current $H_{chiral,0}^\mu$,
in the limit of neglecting all terms of higher order than $O(P^2)$.

If we want $H^\mu$ to reduce to the generalized chiral limit of Eqn.\
(\ref{eqn5}), we have to modify the $q_\mu q_\nu$ term in $\Gamma_{\mu\nu}$:
\be
   \Gamma_{\mu\nu} \to g_{\mu\nu} - \frac{q_\mu q_\nu}{Q^2 - m_\pi^2}
   \approx g_{\mu \nu} - \frac{q_\mu q_\nu}{Q^2} \left( 1 +
   \frac{m_\pi^2}{Q^2} \right)
\ee
and add a non-resonant contribution proportional to $X^{(123)}$ by hand.
The hadronic current for the three pion decay mode then becomes:
\begin{eqnarray}
    H^\mu & = & C^{(3\pi)} \Bigg\{ BW_{a_1}(Q^2)\; \left(g^{\mu\nu} -
    \frac{q^\mu q^\nu}{m_{a_1}^2} \right) \;
   \Bigg( g_{\nu\alpha} - \frac{ q_\nu q_\alpha}{Q^2 - m_\pi^2} \Bigg)
   BW_{\rho}(s_2)\;
\nonumber \\
  & & \times \left(g^{\alpha \beta} -
   \frac{(k_1 + k_3)^\alpha (k_1 + k_3)^\beta}
   {m_\rho^2} \right) \;
   (k_1 - k_3)_\beta
\nonumber \\
   & & + (1 \leftrightarrow 2)
   + \frac{1}{2} \frac{2 m_\pi^2}{Q^2 - m_{\pi}^2} q^\mu
   \Bigg\}
\nonumber \\
   & = & C^{(3\pi)} \Bigg\{ BW_{a_1}(Q^2) BW_\rho(s_2)
      \left[ (k_1 - k_3)^\mu - \frac{(k_1 - k_3) \cdot q}{Q^2 - m_\pi^2}
     \frac{(m_{a_1}^2 - m_\pi^2)}{m_{a_1}} q^\mu \right]
\nonumber \\
   & & + (1 \leftrightarrow 2)
   + \frac{m_\pi^2}{Q^2 - m_{\pi}^2} q^\mu
   \Bigg\}
\end{eqnarray}
Allowing for a  $\rho'$ resonance, we define
\be
   T_\rho(s) = \frac{1}{1 + \beta}\Big\{ BW_\rho(s)
   + \beta \BW_{\rho'}(s) \Big\}
   \label{eqnbeta}
\ee
(see Ref. \cite{ttp9225}, Eqn.\ (20)), and get the final result for
the form factors:
\begin{eqnarray}
   F_1^{(3\pi)} & = & C^{(3\pi)} BW_{a_1}(Q^2) T_\rho(s_2) \nonumber \\
   F_2^{(3\pi)} & = & C^{(3\pi)} BW_{a_1}(Q^2) T_\rho(s_1) \nonumber \\
   F_S^{(3\pi)} & = & C^{(3\pi)} \frac{m_\pi^2}{Q^2 - m_\pi^2}
   \Bigg\{ \frac{Q^2 - m_{a_1}^2}{m_{a_1}^2 Q^2} BW_{a_1}(Q^2)
\\ \nonumber
& &  \mbox{\ \ \ \ \ \ \ \ \ \ \ \ \ \ \ \ \ \ \ }
  \times \left[ \frac{s_3 - s_1}{2} T_\rho(s_2) + \frac{s_3 - s_2}{2}
    T_\rho(s_1) \right] + 1 \Bigg\}
\end{eqnarray}

Now we want to generalize our results to the case of different
pseudoscalars with $m_j^2 \neq m_k^2$. We have to modify the axialvector(A)-%
vector(V)-pseudoscalar(P) coupling in the following way:
\be
   \Gamma_{\mu\nu}\Big(A(q_\mu) \to V(k_\nu) P \Big)
   \to g_{\mu\nu} - \frac{q_\mu q_\nu}{Q^2 - m_{(123)}^2}
   + \frac{1}{2} \frac{q_\mu k_\nu}{Q^2 - m_{(123)}^2}
\ee
If we have a three-particle axial resonance $A$ and two two-particle
vector resonances $V_{13}$ and $V_{23}$ in the $s_2$ and the $s_1$ channels,
respectively,
the hadronic current is given by:
\begin{eqnarray}
   H^\mu & = & C^{(123)} BW_A(Q^2) BW_{V_{13}}(s_2)
   \Bigg\{ (k_1 - k_3)^\mu - q^\mu \frac{q \cdot (k_1 - k_3) (m_A^2 -
   m_{(123)}^2)}{m_A^2 (Q^2 - m_{(123)}^2)}
\nonumber \\
   & & - \frac{m_1^2 - m_3^2}{m_{V_{13}}^2} \Bigg[ (k_1 + k_3)^\mu
   - \frac{q^\mu}{m_A^2 (Q^2 - m_{(123)}^2)}
\nonumber \\
  & & \times \Bigg( q \cdot (k_1 + k_3)
   (m_A^2 - m_{(123)}^2) + \frac{1}{2} (s_2 - m_{V_{13}}^2)
   (Q^2 - m_A^2) \Bigg) \Bigg] \Bigg\}
\nonumber \\
   & & + (1 \leftrightarrow 2)
   + C^{(123)} \frac{1}{2} \frac{X^{(123)}}{Q^2 - m_{(123)}^2} q^\mu
\end{eqnarray}
The normalization is obtained from the chiral limit:
\be
     C^{(123)} = \frac{2 \sqrt{2}}{3 f_\pi} A^{(123)}
\ee
If we allow for two resonances $V_{13}$ and $V'_{13}$ in $s_2$ with
a relative strength defined by $\beta_{13}$ (equivalent to
Eqn.\ (\ref{eqnbeta})) and similarly for $V_{23}$ and $V'_{23}$ with
$\beta_{23}$, we get the general formulae for the form factors:
\begin{eqnarray}
   F_1^{(123)} & = & C^{(123)} BW_A(Q^2) \Bigg\{ \frac{BW_{V_{13}}(s_2)}
   {1 + \beta_{13}}
   \left( 1 - \frac{1}{3} \frac{m_1^2 - m_3^2}{m_{V_{13}}^2} \right)
\nonumber \\ && \mbox{\hspace*{0.6cm}}
   + \frac{\beta_{13}}{1 + \beta_{13}} BW_{V'_{13}}(s_2)
   \left( 1 - \frac{1}{3} \frac{m_1^2 - m_3^2}{m_{V'_{13}}^2} \right)
\nonumber \\ && \mbox{\hspace*{0.6cm}}
   + \frac{2}{3}(m_2^2 - m_3^2)
   \left[ \frac{1}{1 + \beta_{23}} \frac{BW_{V_{23}}(s_1)}{m_{V_{23}}^2}
   + \frac{\beta_{23}}{1 + \beta_{23}} \frac{BW_{V'_{23}}(s_1)}{m_{V'_{23}}^2}
   \right] \Bigg\}
\nonumber \\
   F_2^{(123)} & = & C^{(123)} BW_A(Q^2) \Bigg\{ \frac{BW_{V_{23}}(s_1)}
   {1 + \beta_{23}}
   \left( 1 - \frac{1}{3} \frac{m_2^2 - m_3^2}{m_{V_{23}}^2} \right)
\nonumber \\ && \mbox{\hspace*{0.6cm}}
   + \frac{\beta_{23}}{1 + \beta_{23}} BW_{V'_{23}}(s_1)
   \left( 1 - \frac{1}{3} \frac{m_2^2 - m_3^2}{m_{V'_{23}}^2} \right)
\nonumber \\ && \mbox{\hspace*{0.6cm}}
   + \frac{2}{3}(m_1^2 - m_3^2)
   \left[ \frac{1}{1 + \beta_{13}} \frac{BW_{V_{13}}(s_2)}{m_{V_{13}}^2}
   + \frac{\beta_{13}}{1 + \beta_{13}} \frac{BW_{V'_{13}}(s_1)}{m_{V'_{13}}^2}
   \right] \Bigg\}
\nonumber \\
   F_S^{(123)} & = & \frac{C^{(123)}}{2 (Q^2 - m_{(123)}^2)} \Bigg\{ X^{(123)}
   + \frac{BW_A(Q^2) (Q^2 - m_A^2)}{m_A^2 Q^2}
\nonumber \\ && \mbox{\hspace*{0.6cm}}
  \times \Bigg[\frac{1}{1+\beta_{13}} BW_{V_{13}}(s_2) \Bigg(
   m_{(123)}^2 (Q^2 - 2 s_1 - s_2 + 2 m_1^2 + m_2^2)
\nonumber \\ && \mbox{\hspace*{0.6cm}}
   - \frac{m_1^2 - m_3^2}{m_{V_{13}}^2}
   [m_{(123)}^2 (Q^2 + s_2 - m_2^2) - Q^2( s_2 - m_{V_{13}}^2)]\Bigg)
\nonumber \\ && \mbox{\hspace*{0.6cm}}
   + \frac{\beta_{13}}{1 + \beta_{13}} BW_{V'_{13}}(s_2)
    \times \Bigg( V_{13} \to V'_{13} \Bigg) + (1 \leftrightarrow 2)
   \Bigg]
   \Bigg\}
\end{eqnarray}
For the channels where $G^{(123)} = 0$, the Breit-Wigner resonance
factors $\BW_{V_{13}}$ and $\BW_{V'_{13}}$ must be put equal to zero.

Note that in the case of exact $SU(3)$ flavour symmetry we always have
$m_1^2 = m_2^2 = m_3^2$, in which case the form factors $F_1$ and $F_2$
retain the form they have in the case of exact $SU(3)_L \otimes SU(3)_R$
chiral symmetry. This is of course also true for the three pion decay mode
with three equal masses. The scalar form factor $F_S$, on the other hand,
is always non-zero once the full chiral symmetry is broken, whether or not
the flavour symmetry still holds. $F_S$ gets a non-resonant
contribution which is proportional to pseudoscalar masses squared ($X^{(123)}$)
and a resonant contribution which is proportional to the off-shellness
($Q^2 - m_A^2$) of the axial three particle resonance.

We will now apply these formulae to the channels $K^- \pi^- \pi^+$
and $K^+ \pi^+ K^-$, taking into account $\rho$ and $\rho'$ resonances
in $\pi^+ \pi-$  and $K^+ K^-$ with relative strength $\beta$,
as in Eqn.\ (\ref{eqnbeta}) and a single $K^\star$ resonance in $K
\pi$.
The relevant equations are:
\begin{eqnarray}
   F_1^{(K 2\pi)} & = & - \frac{\sqrt{2}\sin \theta_C}{3 f_\pi}
   BW_{K_1}(Q^2) BW_{K^\star}(s_2) \left(1 - \frac{1}{3} \frac{m_K^2
   - m_\pi^2}{m_{K^\star}^2} \right)
\nonumber \\
   F_2^{(K 2\pi)} & = & - \frac{\sqrt{2}\sin \theta_C}{3 f_\pi}
   BW_{K_1}(Q^2) \left\{
      T_\rho(s_1) + \frac{2}{3} \frac{m_K^2 - m_\pi^2}{m_{K^\star}^2}
   BW_{K^\star}(s_2) \right\}
\nonumber \\
   F_S^{(K 2\pi)} & = & - \frac{\sqrt{2}\sin \theta_C}{3 f_\pi}
   \frac{1}{2(Q^2 - m_K^2)} \Bigg\{(m_\pi^2 + m_K^2)
   + \frac{BW_{K_1}(Q^2) (Q^2 - m_{K_1}^2)}{m_{K_1}^2 Q^2}
\nonumber \\ && \mbox{\hspace*{0.6cm}}
   \times \Bigg[ BW_{K^\star}(s_2) \Bigg( m_K^2 (Q^2 - 2 s_1 - s_2 +
   2 m_K^2 + m_\pi^2)
\nonumber \\ && \mbox{\hspace*{0.6cm}}
   - \frac{m_K^2 - m_\pi^2}{m_{K^\star}^2}
   [m_K^2(Q^2 + s_2 - m_\pi^2) - Q^2 (s_2 - m_{K^\star}^2)] \Bigg)
\nonumber \\ && \mbox{\hspace*{0.6cm}}
   + T_\rho(s_1) m_K^2(Q^2 - 2 s_2 - s_1 + 2 m_\pi^2 + m_K^2)\Bigg]\Bigg\}
\end{eqnarray}
and
\begin{eqnarray}
   F_1^{(K\pi K)} & = & -\frac{\sqrt{2}\cos \theta_C}{3 f_\pi}
   BW_{a_1}(Q^2)\Bigg\{ T_\rho(s_2) + \frac{2}{3} \frac{m_\pi^2 - m_K^2}
   {m_{K^\star}^2} BW_{K^\star}(s_1) \Bigg\}
\nonumber \\
   F_2^{(K\pi K)} & = & -\frac{\sqrt{2}\cos \theta_C}{3 f_\pi}
   BW_{a_1}(Q^2) BW_{K^\star}(s_1) \left(1 - \frac{1}{3}
   \frac{m_\pi^2 - m_K^2}{m_{K^\star}} \right)
\nonumber \\
   F_S^{(K\pi K)} & = & -\frac{\sqrt{2}\cos \theta_C}{3 f_\pi}
   \frac{1}{2 (Q^2 - m_\pi^2)}\Bigg\{ (m_\pi^2 + m_K^2) +
   \frac{BW_{a_1}(Q^2)(Q^2-m_{a_1}^2)}{m_{a_1}^2 Q^2}
\nonumber \\ && \mbox{\hspace*{0.6cm}}
   \times \Bigg[ T_\rho(s_2) m_\pi^2 (Q^2 - 2s_1 - s_2 + 2 m_K^2 + m_\pi^2)
\nonumber \\ && \mbox{\hspace*{0.6cm}}
   + BW_{K^\star}(s_1)\Bigg(m_\pi^2 (Q^2 - 2 s_2 - s_1 + 2 m_\pi^2 +
   m_K^2)
\nonumber \\ && \mbox{\hspace*{0.6cm}}
   - \frac{m_\pi^2 - m_K^2}{m_{K^\star}^2}
   [m_\pi^2 (Q^2 + s_1 - m_K^2) - Q^2 (s_1 - m_{K^\star})]\Bigg)\Bigg]
   \Bigg\}
\end{eqnarray}
Note that the anomalous form factor $F_3$ is not affected by the pseudoscalar
masses, since the anomaly is a short distance effect.
Therefore in the numerical discussion in the next section we use
the same anomalous form factors as in Ref.\ \cite{ttp9225}.

Finally, we do not consider scalar resonances in this paper. For the inclusion
of a $J^P = 0^-$ resonance ($\pi'$) and its possible effects we refer to
Refs. \cite{ttp9225,KM2}.
\section{Numerical Results}
We will start the numerical discussion by giving the integrated decay rates
$\Gamma^{(abc)}$,
normalized in the usual way to the electronic width $\Gamma_e$ of the
tau. The total width gets three contributions (cf. Eqns. (\ref{eqnrate1})--%
(\ref{eqnrate2}) and (\ref{ws})):
\be
   \Gamma^{(abc)} = \Gamma_n^{(abc)} + \Gamma_a^{(abc)} + \Gamma_S^{(abc)}
\ee
$\Gamma_n^{(abc)}$ is the ``normal'' contribution resulting from the
form factors $F_1$ and $F_2$,
$\Gamma_a^{(abc)}$ is the anomalous contribution
resulting from $F_3$, and $\Gamma_S^{(abc)}$ is the scalar contribution from
$F_S$. For the $3\pi$ channel, $\tau^-\rightarrow\nu  \pi^-\pi^-\pi^-$,
we use the parametrization of \cite{ttp9225} (see also \cite{KM2})
for the Breit-Wigners $BW_{a_1}$ and $T_{\rho}$.
In the case of the
channel $\tau^-\rightarrow\nu K^-\pi^-\pi^+$ ,
the parametrizations of the Breit-Wigner factors $BW_{K^*}$, $BW_{K_1}$ and
$T_{\rho}(s_i)$ are taken from \cite{DMREV},
where Eqn.\ (35) in \cite{DMREV} is used for the $T_{K^*}$ Breit-Wigner
in the three body resonance, which occurs in the form factor $F_3$.
The other parameters can also be found in \cite{ttp9225}.
The parametrization for the decay $\tau\rightarrow\nu K^-\pi^-K^+$
is also taken from \cite{ttp9225} and \cite{DMREV}.

With these parametrizations we get the following results:
\begin{center} \begin{tabular}{ccccc}
Channel $(abc)$ &
$\frac{\Gamma_{}^{(abc)}}{\Gamma_e}$ &
$\frac{\Gamma_{n}^{(abc)}}{\Gamma_e}$ &
$\frac{\Gamma_{a}^{(abc)}}{\Gamma_e}$ &
$\frac{\Gamma_{S}^{(abc)}}{\Gamma_e}$ \\ \\
$\pi^- \pi^- \pi^+$ & 0.356 & 0.356 & 0 & 0.0000073 \\
$K^- \pi^- \pi^+$   & 0.0327 & 0.0313 & 0.00137 & 0.0000033 \\
$K^- \pi^- K^+$     & 0.0061 & 0.0037 & 0.0023 & 0.0000013 \\
\end{tabular} \end{center}
We find that the relative contribution of the scalar part is  of the order of
$10^{-5}$ in the $3\pi$ case,
and $10^{-4}$  in the channels $K^- \pi^- \pi^+$
and   $K^- \pi^- K^+$.
Note that especially in the modes with kaons our results
for the scalar part are actually much smaller than the naive application of
the PCAC argument would indicate (cf. the estimates in the introduction).
So we find that in no case the scalar contribution could be measured
in the total decay width.
In the modes with mesons of different masses (ie.
$K^- \pi^- \pi^+$ and $K^- \pi^- K^+$), the form factors $F_1$ and $F_2$ are
also modified by the inclusion of the pseudoscalar mass effects, but the
numerical size of this effect is negligible (less than 1\%).
Considering the large uncertainties in the predictions for the rate which
result from details of the Breit-Wigner parametrizations \cite{ttp9225},
it is clear that also these effects on the total rate can not be used to see
the pseudoscalar mass effects experimentally.

So we have to consider angular distributions as suitable
observables. We will therefore
present now numerical results for the
spin-0-spin-1 interference structure
functions $W_{SB,SD,SF}$ and the pure spin-0 structure function $W_{SA}$
and compare them with the pure spin-1 structure functions $W_{E,G}$ and
$W_I$. In particular we will concentrate on the $Q^2$ distribution
of the structure functions, i.e. we integrate over the Dalitz-plot
variables $s_1$
and $s_2$.
Note that the most interesting moments for our analysis
in Eqn.\ (\ref{moments}) projects only on a linearcombination of
one spin-1 and  one  spin-0 structure function.

There are two possible effects contributing to the scalar form factor: The
pseudoscalar masses, considered in the present paper, and scalar ($J^P = 0^-$)
three particle resonances \cite{ttp9225,KM2}.
The $Q^2$ dependence of the structure functions  for a possible $J^P=0^-$
 resonance
is already discussed in \cite{KM2}.
As we will show below
the $Q^2$ dependence of the
impacts on the structure functions of these
different effects are very different: The scalar resonance contribution
is peaked around the resonance mass, whereas the pseudoscalar mass effects
are large at low $Q^2$. So by measuring the $Q^2$ distributions these two
effects can be distinguished.

Let us start with the $3\pi$ channel: $\tau^-\rightarrow\nu  \pi^-\pi^-\pi^-$.
As already mentioned, due to G-parity conservation $W_G$ and $W_I$ are
vanishing
 in
this case and a nonvanishing contribution to the moments
$\langle \sin\beta \cos\gamma \rangle$ and
$\langle \sin\beta \sin\gamma \rangle$ yields a clean signature of the presence
 of
a scalar contribution and a measurement would allow to analyse the scalar
 form factor
$F_s$ in detail.
In Fig.\ 2a we show the $Q^2$ distribution of the $s_1,s_2$ integrated
structure
 functions
$W_{SA,SB,SD}$ normalized to
 $W_{tot}=W_A+W_{SA}$. Note that $W_B$ vanishs in the
three pion channel.
One observes a sizable contribution of the scalar form factor only at low $Q^2$
values ($Q^2<0.8 GeV^2$).
 This is in contrast to a possible scalar resonance contribution
to the scalar form factor where the scalar resonance is peaked around the
$\pi'$
resonance mass, see Fig.\ 4 in \cite{KM2}.
A measurement of the $Q^2$ dependence would therefore allow to disentangle
these two possible contributions.
It is clear that the  scalar form factor effect is strongly enhanced by the
 interference
with the larger spin-1 form factors $F_{1,2,3}$, whereas the pure spin-0
structure function remains small over the whole $Q^2$ range.
For comparison, we show the normalized spin-1 structure function $W_E/W_{tot}$
 in Fig.\ 2b
 as a function
\footnote{In the three
 pion case the moment $\cos\beta$ is combined
with an energy ordering $sign(s_2-s_2)$ to account
 for Bose Symmetry, see \cite{KM1,KM2}.}
of $Q^2$.
 As mentioned before this ratio is closely related
to the parity violating asymmetry \cite{KM1,KM2}. Note that in contrast to
the figures in \cite{KM1,KM2}, we have taken  the spin-0 contribution in
the normalization $W_{tot}$ into account.

Let us now discuss the numerical effect of the nonvanishing meson masses to the
Cabibbo suppressed $\tau\rightarrow\nu
K^-\pi^-\pi^+$ channel.
In this case there are contributions to all structure functions in Eqn.\
 (\ref{ws}).
Fig.\ 3a shows the $Q^2$ dependence of the $s_1,s_2$ integrated structure
 functions
$W_{SA,SB,SD,SF}$ normalized to $W_{tot}=W_A+W_B+W_{SA}$. Like in the three
pion
 case
a sizable contribution of the scalar form factor is only observable at low
$Q^2$ values and the pure spin-0 contribution remains small over the whole
$Q^2$
range.
The results for the normalized spin-1 structure functions are shown in Fig.\
3b.
(A detailed discussion of the latter structure functions can also be found in
\cite{DMREV}.)
We find that
the spin-1 structure function $W_{I}$ is rather small in the region where
the spin-0-spin-1 interference structure function $W_{SB}$ becomes large,
so the scalar part can indeed be measured.

Finally we present results for the Cabibbo allowed decay $\tau\rightarrow\nu
 K^-\pi^-K^+$.
In Fig.\ 4a we show the results for the normalized scalar structure functions
$W_{SA,SB,SD,SF}$ again nomalized to $W_{tot}=W_A+W_B+W_{SA}$.
For $Q^2$ values below 1.6 $GeV^2$ the mass effects are fairly large.
For comparison, predictions for the pure spin-1 structure functions are shown
in
 Fig.\ 4b.
(The latter have also been discussed in \cite{DMREV}.) We find again
that in the important low $Q^2$ region the spin-0-spin-1 interference
structure functions are comparable with the corresponding pure spin-1
structure functions.

\section{Conclusions}
We have shown how the well-known approach
of extrapolating from the chiral limit
to higher energies by Breit-Wigner resonances can be generalized
by extrapolating from massive rather than massless pseudoscalar
mesons. The inclusion of the pseudoscalar  masses does not
change the predictions for the integrated decay rate significantly.
Nevertheless the non-conservation of the axial current leads to a
non-vanishing scalar form factor which can be measured in angular distributions
by suitable spin-0-spin-1 interference effects and can  be
distinguished clearly from a pseudoscalar resonance by the $Q^2$ distribution.
Therefore we now urge for a careful experimental analysis of the
scalar form factor in these tau decays, which would enhance our understanding
of the structure of the hadronic current and which would also be important
for other analyses which
make certain assumptions about the scalar form factor (eg. the measurement
of the tau polarization by using the three pion decay).

\section*{Figure captions}
\begin{description}
\item[{Fig.\ 1}]
 diagrams for the vector meson dominance model of the  decay $\tau^- \to
\pi^- \pi^- \pi^+$.
\item[{Fig.\ 2}]
 $Q^{2}$ dependence of  $s_1$, $s_2$ integrated structure functions  for the\\
decay channel $\tau\rightarrow \nu\pi^-\pi^{-}\pi^{+}$:\\
a)    $W_{SA},W_{SB},W_{SD}$ (solid, dashed, dotted) normalized to
    $W_{tot}$.  \\
b)    $W_E$ normalized to $W_{tot}$.
\item[{Fig.\ 3}]
 $Q^{2}$ dependence of  $s_1$, $s_2$ integrated structure functions  for the\\
decay channel $\tau\rightarrow \nu K^-\pi^{-}\pi^{+}$:\\
a)    $W_{SA}, W_{SB}, W_{SD}, W_{SF}$ (solid, dashed, dotted, dashed-dotted)\\
       normalized to     $W_{tot}$.  \\
b)    $W_E, W_G, W_I$ (solid, dashed, dotted) normalized to $W_{tot}$.
\item[{Fig.\ 4a,b}]
same as Fig.\ 3 for the decay channel $\tau\rightarrow \nu K^-\pi^{-}K^{+}$.
\end{description}
\end{document}